# A New System Architecture for Pervasive Computing

Anis ISMAIL[1], Abd El Salam AL HAJJAR[1] and Ziad Ismail[2]

[1]Institut Universitaire de Technologie
Université Libanaise
Saida, BP 813, LIBAN
anismaiil@ul.edu.lb
abdsalamhajjar@hotmail.com

[2] TELECOM ParisTech
Paris, France
ismail.ziad@enst.fr

## ABSTRACT

*We present new system architecture, a distributed framework designed to support pervasive computing applications. We propose a new architecture consisting of a search engine and peripheral clients that addresses issues in scalability, data sharing, data transformation and inherent platform heterogeneity. Key features of our application are a type-aware data transport that is capable of extract data, and present data through handheld devices (PDA (personal digital assistant), mobiles, etc). Pervasive computing uses web technology, portable devices, wireless communications and nomadic or ubiquitous computing systems. The web and the simple standard HTTP protocol that it is based on, facilitate this kind of ubiquitous access. This can be implemented on a variety of devices - PDAs, laptops, information appliances such as digital cameras and printers. Mobile users get transparent access to resources outside their current environment. We discuss our system's architecture and its implementation. Through experimental study, we show reasonable performance and adaptation for our system's implementation for the mobile devices.*

## 1. INTRODUCTION

The use of computers is increasing in every aspect of our lives, in shopping, moving from one place to another, in communication. The advances in computing and communication over wired and wireless networks have resulted in many pervasive computing environments. There are some examples The Internet, intranets, local area networks, mobile ad hoc wireless networks, peer-to-peer networks, and sensor networks. Technology appears clearly in the way we live. Definitely, we are going directly toward pervasive computing.

Pervasive computing permits to users to use their computers to accomplish tasks anytime-anywhere (7days*24 hours), with any device (PC, Palm/PDA, Cell phone, TV…), from any network that enable you to accessing and to notifying any data ( such as E-mail, Personal Information, public/private Services), to make data synchronization and transaction, to use secure connection using wireless optimization. Therefore, the objective of Pervasive computing is to permit any authorized user to access any data and any available application services anywhere, anytime with any devices.

"The Internet revolution has barely started. It won't be done until everything is on the Web. Pagers, light switches, printers, copiers, as well as PCs, benefit from Web connections." [1]

In the future, it is not only by the desktop computers, you can access the Web, but also there are others equipments, appliances in the home, which are also connected to the internet like refrigerators, oven, Light switches, cameras, microwave, etc. In a living room, you can use a customized TV remote equipped with a TrackPoint pointing stick to bring up a Web page on a large-screen television. And also you can switch to a wireless WebDT digital tablet, which lets him control almost any fixture in the house, including the speed of the kitchen-ceiling fan, temperature of the refrigerator with audio system warning or by sending information to your PDA or mobile inform you what were missing, and the lighting level in the living room. Consequently, the world tends to mobility in its daily life activities, in the direction of ubiquitous computing.

The concept of ubiquitous computing began, not as an exercise in using the Internet, but as a study of how people work, how they use tools and where the future of computing might lie. The objective of ubiquitous computing is to move computers away from the central focus of the user's attention into the invisible world, where they are used subconsciously, to enhance existing tools or communications, and to make the user free from time and space constraints. So, the goal of ubiquitous computing is the production of devices that are so commonplace and natural to use that they become almost invisible. The devices being ubiquitous means there will be hundreds of small computers in an office or home, each doing its own specialized task. Ubiquitous Data Mining is one of the potential applications that motivate the need for ubiquitous computing and is concerned with data analysis and delivery on mobile devices. Mining the Internet environments naturally calls for proper utilization of the ubiquitous resources. Ubiquitous Data Mining offers an alternate approach to address this problem of mining data using ubiquitous resources and pays careful attention to the ubiquitous resources of data, computing, communication, and human factors in order to use them in a near optimal fashion.

Ubiquitous computing puts computing in the border of our lives, as a tool, not a focus, out of the way so that we can get on with the true tasks we wish to do. The term 'ubiquitous' is used because computers and computation will be everywhere, embedded in the fabric of our lives.

An example of a ubiquitous technology present today is literacy. Words are displayed on every surface and body part imaginable, and they convey information to us automatically, without invoking our conscious mind. We don't suffer from 'literacy overload', cases of stress because we are being surrounded by too many words (unless it is that stack of journals every engineer feels compelled to read). We automatically read a street sign and it guides us without undue effort on our part. Ubiquitous computing will bring the Internet into our daily lives with less effort. Instead of keeping lists of pertinent URLS or 'favorite places' on our browsers, the devices that need the information can find it themselves. Instead of 'surfing' to find and sift through all of the information available to us, some other agent or device will do the searching for us.

In this paper, we present a new application, a distributed framework designed to support pervasive computing applications. We propose a new architecture consisting of a search engine and peripheral clients. Key features of our application are a type-aware data transport that is capable of extract data, and present data through handheld devices (PDA (personal digital assistant), mobiles, etc).

The remainder of this paper is organized as follows. Section 2 offers an overview of the literature on Pervasive computing system. Our system architecture in pervasive issues is described in the section 3. In the section 4, we represent a new web application applicant to our architecture.. In the last section, we conclude the main results of this paper.

## 2. RELATED WORK

Pervasive Computing strives to simplify daily life by providing the means of carrying out personal and business tasks via portable and embedded devices. These tasks could be as simple as switching on the lights in a conference room, checking email, organizing meetings, accessing services in a room, to booking airline tickets, buying and selling stock, and even managing bank accounts.

As part of our research into pervasive computing that has a objective of providing "anytime, anywhere, anyplace" computing by decoupling users from devices and viewing applications as entities that perform tasks on behalf of users [2] is built "one.world" a system architecture for pervasive computing. [3] developed Centaurus which realizes the Smart Office scenario, where intelligent services are accessible to mobile users via hand-held devices connected over short range wireless links.

Also, there are numerous ongoing projects in this area, including PIMA [4] that is founded on the idea that the application model for pervasive computing must decouple application logic from details that are specific to the run-time environment, such as specific services and user interface renderings. Application functionality is modeled in a generic fashion as tasks and sub-tasks joined together by navigation mechanisms, Aura and Portolano [5].

The Aura pervasive computing project is the successor to the earlier Coda and Odyssey projects on adaptation. Like PIMA, Aura proposes a programming model for task-based computing [6] In this model, tasks are viewed as compositions of services. Both tasks and services have explicit representations. Services are described by virtual service types, which define functional, state and configuration interfaces and dependencies upon other services. Virtual service types can be related through inheritance, and can also be composed to form new virtual services. Tasks are top level compositions of services that are specified as flows that decompose tasks into steps of subtasks or primitives (actions carried out by services)

The Portolano project [5], in contrast to PIMA and Aura, primarily addresses issues of infrastructure rather than of software development. The Portolano group proposes the use of data-centric networks, an approach based on active networks in which data packets are responsible for traversing the network and obtaining required resources inside the network. The group is also considering infrastructural issues such as service discovery and proxy architectures that support resource poor devices, and has an interest in applications such as location tracking of objects, gathering of data from sensors and applications of embedded Web servers. The Portolano research currently remains in its early stages.

[7] describes the research challenges in developing an application model for pervasive computing. The purpose of that paper was to examine basic questions such as the nature of pervasive applications, the role of pervasive devices, and the relevance of a user's environment. [7] axamine basic questions such as the nature of pervasive applications, the role of pervasive devices, and the relevance of a user's environment.

An important type of pervasive computing service is that which exposes an "information interface" to physical artifacts, via sensors and actuators. Much work has been done by other researchers [8] on building sensor frameworks. This is an emerging research area with plenty of open issues.

The Gaia system in [9] [10] aimed at developing a distributed middleware infrastructure that coordinates software entities and heterogeneous networked devices contained in a physical space. The physical space is termed as an active space in which users interact with several devices and services at the same time. Gaia exports services to query access, and uses existing resources and provides a framework to develop user centric mobile applications.

Nevertheless, pervasive computing is faced with important issues that may stand as obstacles in developing these types of systems. The issue of uneven conditioning which we mentioned earlier deals with resource-hungry users who need more resources than what is currently available to them. One solution for such problem was proposed by Project Aura [5]. The solution informs users of a better spot to obtain a service, as well as the resources necessary for the required task. However, such approach requires knowledge of the exact task at hand, along with intelligence about the distribution of resources. Another important aspect which is under intensive research in pervasive computing is the scalability issue. Scalability is recognized as a primary factor that influences the architecture and implementations of pervasive systems, similar to what has been done in distributed systems [11].

Gurumurthy et al. [17] represents challenge for mobile computing community by questioning the roles of devices, applications, and a user's environment. A vision of pervasive computing is described, along with attributes of a new application model that supports this vision to reality. Pervasive computing is more art than science. It will remain this way as long as people continue to view mobile computing devices as mini-desktops, applications as pro-grams that run on these devices, and the environment as a virtual space that a user enters to perform a task and leaves when the task is finished.

Network computing and mobile computing are fast becoming a part of everyday life. [18] expect devices like mobile phones, PDAs, offices PCs and even home entertainment systems to access information and work together in one integrated system and the challenge is to combine these technologies into a seamless whole and on the Internet. The goal of Pervasive Computing is for computing available wherever it's needed. It spreads intelligence and connectivity to more or less everything. So conceptually, ships, aircrafts, cars, bridges, tunnels, machines, refrigerators, door handles, lighting fixtures, hats, shoes, tools, packaging clothing, appliances, homes and even things like our coffee mugs and even the human body and will embedded with chips to connect to an infinite network of other devices and to create an environment where the connectivity of devices is embedded in such a way that it is unobtrusive and always available. Therefore, pervasive computing, refers to the emerging trend toward numerous, easily accessible computing devices connected to an increasingly ubiquitous network infrastructure.

Khan et al. [19] report on the challenges of in-situ evaluation of a pervasive computing application targeting working parents. Apart from technical issues that were also reported in the aforementioned studies, trivial unforeseen problems such as participants having to carry a second mobile device next to their own phone also negatively affect the evaluation efforts.

Although the theoretical benefits of evaluating pervasive computing applications in-situ are substantial, the challenges are still considerable. In some cases while pursuing to overcome those challenges the ecological validity of evaluation studies becomes questionable.

Consolvo et al. [20] discuss the limitations of an in-situ evaluation. Their first study researched a prototype deployed in a home setting for supporting eldercare. The prototype was mimicking sensor data with a Wizard of Oz. In-situ data were collected daily over the phone. Their second study researched a location aware application to understand the link between user preferences and destinations. In this study, situ self-reports via messages triggered based on the participants' arrival at a destination were collected along with interview data. Their third study investigated daily physical activity and whether sharing activity related data with a small group of friends might influence physical activity goals. This study combined interviews and questionnaires with in situ user-initiated logging of pedometer data mobile phone application.

Khan et al. [21] present a setup for evaluating users' experience of pervasive applications within a virtual environment. They review existing literature on mixed reality and pervasive application evaluation. A conclusion of that review is the potential of evaluating applications such as location-based services in a virtual environment. Finally, they present their plans of evaluating user experience factors of location-based advertisements in a virtual supermarket, highlight methodological considerations and sketch future research directions.

Pervasive computing provides an attractive vision for the future of computing. Computational power will be available everywhere. Mobile and stationary devices will dynamically connect and coordinate to seamlessly help people in accomplishing their tasks. For this vision to become a reality, developers must build applications that constantly adapt to a highly dynamic computing environment. To make the developers' task feasible, [22] present a system architecture for pervasive computing, called one.world. Their architecture provides an integrated and comprehensive framework for building pervasive applications. It includes services, such as discovery and migration that help to build applications and directly simplify the task of coping with constant change. They describe their architecture and its programming model and reflect on their own and others' experiences with using it.

[23] discusses how we have adapted and applied traditional methods from psychology and human–computer interaction, such as Wizard of Oz and Experience Sampling, to be more amenable to the in situ evaluations of ubiquitous computing applications, particularly in the early stages of design. Although the focus is on ubiquitous computing applications and tools for their assessment, it is believed that the in situ evaluation tools that are proposed will be generally useful for field trials of other technology, applications, or formative studies that are concerned with collecting data in situ

## 3. SYSTEM ARCHITECTURE

Pervasive Computing is categorized by the interaction of a multitude of heterogeneous devices, ranging from powerful general-purpose servers located in an infrastructure to tiny mobile sensors integrated in everyday objects. Devices are connected to each other using wireless communication technologies like Bluetooth, IrDA or IEEE 802.11. The devices that are present in a certain environment share their functionality with other devices in the vicinity to win mutual benefits. A sensor can for instance use a display to present its data to a person nearby.

The development of applications for dynamic environments is a non-trivial job. Induced by mobility, fluctuating network connectivity or changing physical context, the software and hardware resources available to an application at runtime are continuously fluctuating. As a result, applications need to adapt themselves continuously to their ever-changing execution environment.

The principal components of our architecture (Figure1) are the devices, the clients (users), the web pages, the Web Services (UDDI) and the ubiquitous sources.

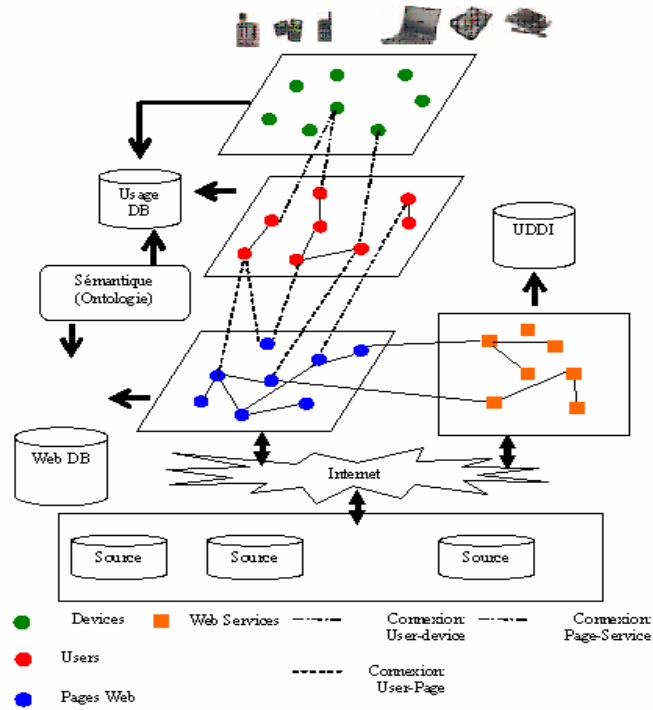

Figure 1. System Architecture

Nowadays, when we want to access the WWW (World Wide Web), we find ourselves standing in front of Personal computers which are groups of a NC (Network Computer), and which are sending and receiving so much of data. These devices force the user to have minimum technical knowledge to access the internet to use new media including Web sites, streaming audio and video, chat rooms, e-mail, online communities, Web advertising, DVD and CD-ROM media, virtual reality environments, etc. There are all sorts of devices other than computers connected to the Web, Laptops, telephones, printers, robot arms, cameras, pagers, Handheld devices (PDA (personal digital assistant), mobiles, etc). The unpredictable growth of wireless access to the Internet gives you the opportunity to check your email and favorite Web sites while you're on the way to your job or to anywhere too. The clients use these devices (cell phone, PDA…) with advanced technologies, to check their E-mail. And the developers are going through this trend by creating Web sites that are customized for handheld devices that contain. These include everything on the internet that deliver news headlines, stock quotes, and other frequently updated information, to shopping sites.

The connection between the devices and the internet are varied, upon their deployment around the world. The most widely used are wireless Wide Area Networks, Satellite and GSM technology.

These devices have to contain several components necessary to facilitate the creation of secure and scalable pervasive devices. . These are Application Components (Service Management Framework and Device libraries, Java Virtual Machine (JVM), Signature Applications (Speech Recognition), Build in HTTP server, and Data management (DBMS)), User Interface Components (widgets, browsers, text to speech and speech recognition components (voiceXML), and plug-ins for multimedia), Communication Components (TCP/IP, Bluetooth, WAP, Wi-Fi (802.11)), Application Development Tools (standard java development tools), and Security Components (Cryptography, SmartCards, Secure Sockets Layer (SSL)).

Some devices can be connected to other devices which are connecting to the internet, through a several remote control services, for example remote desktop sharing, remote administrator, etc.

The clients can access favorite web sites by giving the device a URL and can bookmark the URL to use later, or downloads Web pages (more precisely, HTML-based data) and lets you browse them offline on your Handheld device, so they are easily accessible, and they could be connected to another web sites to emphasize the meaning of that web sites contents, heterogeneous data sources, and also we can use services embedded in these web sites, and these services could be connected to other services to accomplish the goal and also can use the services listing in UDDI (Universal Description, Discovery and Integration) that is one of the core Web services standards [12] and is an open industry initiative, enabling businesses to publish service listings and discover each other and define how the services or software applications interact over the Internet.

## 4. RECOMMENDATION ENGINE SPONSORED BY THE AMAZON APPLIED TO OUR ARCHITECTURE

On the Internet, recommendation tools specially equipped merchant sites, such as Amazon. The manual is known: you choose the title of a book, and a tool recommends you to other secure sites, recorded by Internet users. suggestion tools are best known for their use on e-commerce Web sites [15], where they use input about a customer's interests to generate a list of recommended items. Multiple applications use only the items that customers purchase and explicitly rate to represent their interests, but they can also use other attributes, including items viewed, demographic data, subject interests, and favorite artists.

Jean-Baptiste Rudelle said: "The catalogues of merchants becoming too important. Entering a keyword can sometimes go back over 5000 results. Therefore, the filtering is absolutely necessary. This is one of the challenges of tomorrow's Internet: filter and present the right information to the right people" [16].

Based on a numerous of mathematical algorithms, the engine recommendation addresses on three types of sites: e-merchants, portals content (sites ads or job offers,…) and blogs . The technology is currently marketed mainly in white mark in Europe: "Some 3000 Web sites have incorporated into their system and we serve 12 million recommendations per day," said Jean-Baptiste Rudelle citing two clients emblematic: site distribution and DVD rental Glowria and the portal specialized AlloCine. A second fundraising in preparation should allow Criteo to increase its presence in the USA.

Our tool "Engine of recommendation sponsored by the Amazon" is a Web application ASP. NET, which allows the user to search for books, software, computer from Web Amazon for any item in their catalogues using Service Web Amazon (version 3.0).

Modeling and understanding the context of a user is a huge open issue that several researchers are working on [13]. The other problem mentioned above is called "Dynamic service discovery". The industry has made progress with service discovery with the development of standards such as UDDI [14].

**4.1. User interface (UI)**

Figure 2. Initial application interface.

The interface (Figure 2) of our engine supported by Amazon is a basic one that seeks to capture all necessary input from the user in an elegant and simple form.

Below we can see screenshots of when we lunched the application and one of the application displaying results Figure 3) from a previous search:

Figure 3. Results page.

In the left of the application, these two screens are what you will come to see. The only exception being if an error occurs during the search, at which time you will see an appropriate error message displayed. The interface contains just a few ASP.NET Web Form controls. we used normal HTML versions of the Web Form controls you see, namely, Drop-Down lists, text boxes, and a submit button.

In Figure 4, you will note the numbers 1-2-3. Each of these areas contains different controls or components that comprise the entire application.

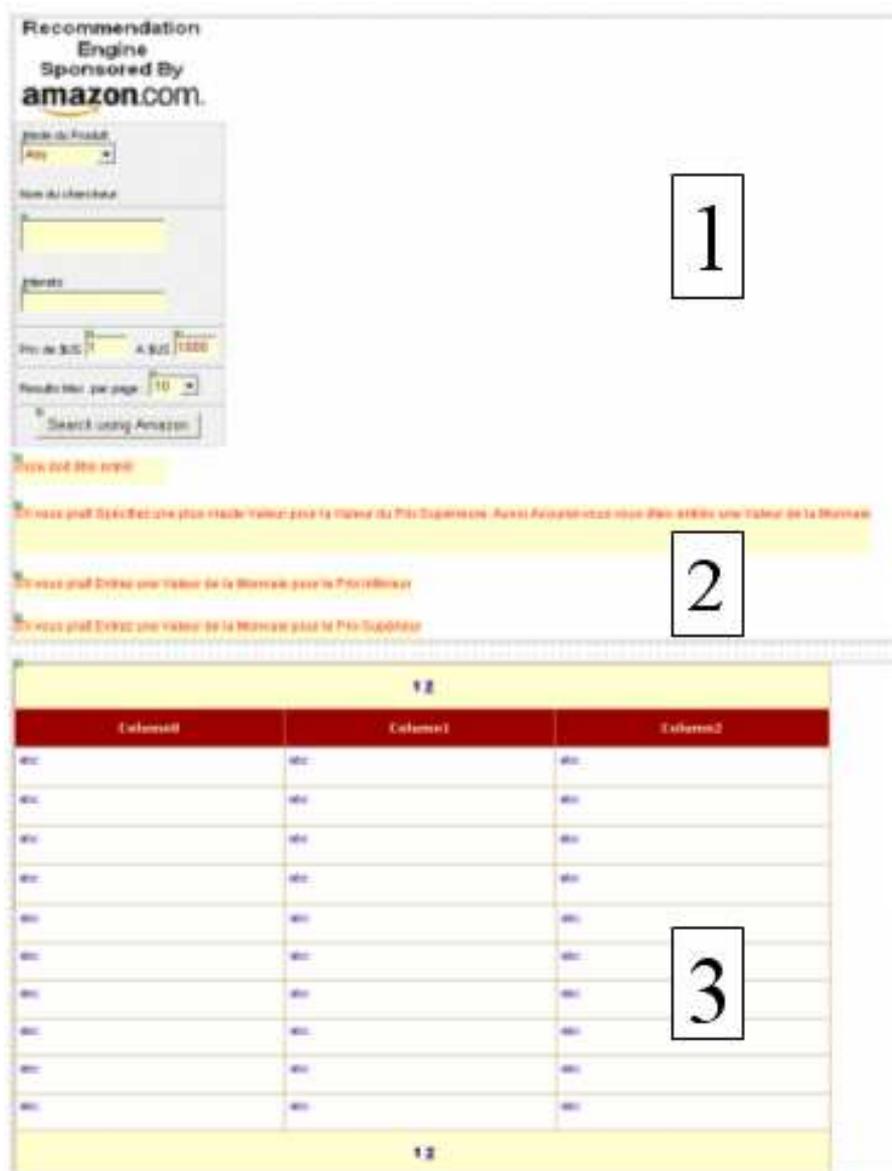

Figure 4.  Visual Studio .NET IDE -- Design mode

Partie 1: Researcher name, interest, Drop-Down lists, text boxes, and submit button

Drop Down lists were used to accommodate the easy selection by users of search parameters that aided in the search for the item in the Amazon catalog. One drop-down list was used allowing the user to select the category they think their item will be found in. The other list allows the user to identify the maximum number of results they want the search to produce on the results page.

Text boxes were used to specify the needed of the users. The other text boxes, we collect price range values so that the user can search for an item in a price range. regrettably the Amazon Web Service does not support price ranges via SOAP but do support it via normal HTTP requests.

Partie 2 : Validator controls

Validator controls used to help validate user input before the form values are used to commit a search against the Amazon Web Service. To perform input validation here, we have used the .NET CompareValidator Web Form control which quite nicely allows you to specify the items to be compared by just choosing them in the properties of the control.

Partie 3: DataGrid component

The DataGrid component used to display the results of a search by the end-user. This component is fairly ideal for the job as it can be bound to various data sources. The data source is used depend on your needs, but for this scenario I used a DataSet. The data source for the search populated via the Amazon Web Service.

### 4.2. Source code

After we have implemented a User interface, we need some code that processes the input from our User Interface to give us some meaningful output by means of displaying some results to the end-user. Our implementation language of choice in this project is C# which gives us the flexibility to implement several applications.

The action starts in our system when the user clicks the "Search using Amazon" button. The Web Form is sent back to the server where the input is gathered and run through the Amazon Web Service which returns some results to the Web Form for display to the user.

### 4.3. Gathering & caching data

The goal of our system is displaying results from an end-user search. There are times when there will be LOTS of results. To solve this issue is to have a mechanism to page the results displaying "x" results at a time.

The DataGrid offer this functionality with the simple selection of a few property fields of the control. The fields involved in this are namely, AllowPaging and PageSize. Note also that we can implement our own custom paging solution for the DataGrid, however, we just wanted simple paging so we went with the property selections. Along with these property selections, one also has to create an event-handler for the PageIndexChanged event that is fired by the DataGrid when a user selects a different page than the current page.

```
1: private void gridAmazonResults_PageIndexChanged
```

```
           (object source, DataGridPageChangedEventArgs e)
2: {
3:    gridAmazonResults.CurrentPageIndex = e.NewPageIndex;
4:    dsetAmazonReponse = (DataSet)Cache[AMAZON_CACHE_NAME];
5:    gridAmazonResults.DataSource = dsetAmazonReponse;
6:    gridAmazonResults.DataBind();
7: }
```

It is important to note that all that is being done is to change the current page index in the DataGrid (line 3) and then bind the DataGrid (line 6) to a cached DataSet (line 4-5).

Caching in ASP.NET is rather simple and shows remarkable speed increases in terms of usability and processing of data.

```
1: private void btnSubmit_Click(object sender, System.EventArgs e)
2: {
3:    Cache.Remove(AMAZON_CACHE_NAME);
4:    dsetAmazonReponse = (DataSet)Cache[AMAZON_CACHE_NAME];
5:
6:    if( dsetAmazonReponse == null )
7:    {
8:       dsetAmazonReponse = new DataSet();
9:       dsetAmazonReponse.Tables.Add("AmazonResults");
10:      dsetAmazonReponse.Tables["AmazonResults"].Columns.Add("Preview");
11:      dsetAmazonReponse.Tables["AmazonResults"].Columns.Add("Name");
12:      dsetAmazonReponse.Tables["AmazonResults"].Columns.Add("Our Price");
13:      dsetAmazonReponse.Tables["AmazonResults"].Columns.Add("List Price");
14:      dsetAmazonReponse.Tables["AmazonResults"].Columns.Add("Catalog");
15:
16:      ProcessSearchRequest();
17:      Cache[AMAZON_CACHE_NAME] = dsetAmazonReponse;
18:   }
19:   dsetAmazonReponse = (DataSet)Cache[AMAZON_CACHE_NAME];
20:   gridAmazonResults.DataSource = dsetAmazonReponse;
21:   gridAmazonResults.DataBind();
22: }
```

In the above code extract we see what's going on when a user requests a "brand new" search. We first remove the current DataSet that is in the cache (line 3-4) then we acquire a new DataSet and create the table we use to setup the data we get from the Amazon Web Service (line 6-18). After the request has performed by the Amazon Web Service we rebind our DataGrid to the DataSet that now has the data we want to display in the grid to the user in a nice formatted manner (line19-21).

#### 4.4. Application configuration

Like most applications, our application has certain characteristics that can be configured to affect how the application runs. One such example is the configuration that allows the application to be run behind a proxy. We wanted a mechanism that would allow us to change a flag or something to that effect to allow our application to work behind a proxy without me having to recompile code and all that fun stuff. Again, the .NET framework came through for us.

Microsoft has a mechanism that now allows developers to put their application specific settings in what they called .config files. Really, these files are quite similar to .ini files with the only difference being that they are in XML format. With this feature, we were able to specify a few settings that would change as the application was distributed to various domains, machines, and developers.

```
<appSettings>
  <add key="SetProxy" value="off" />
  <add key="ProxyIP" value="10.132.0.4" />
  <add key="ProxyPort" value="8080" />
</appSettings>
```

Above we are able to specify in the Web.config file some properties that we consider dynamic in the application. The proxy information allows us to change the proxy information for my location just by editing the .config file. When the user clicks the "Search using Amazon" button, the proxy will be used depending on the value of the SetProxy field in the .config file.

```
1: private void Page_Load(object sender, System.EventArgs e)
2: {
3:   if( ConfigurationSettings.AppSettings[CONFIG_PROXY_NAME] == "on" )
4:     SetProxy(ConfigurationSettings.AppSettings[CONFIG_PROXY_IPVALUE],
5:       Convert.ToInt32(ConfigurationSettings.
6:           AppSettings[CONFIG_PROXY_PORTVALUE]));
7: }
```

In line 3, you can see how we can access the settings specified in the .config file from the Web Form code. The AppSettings collection of the configuration object contains all the key-value fields we specified in the <appSettings> element of the XML based .config file. Gaining access to the values of the settings is a simple matter of indexing into the AppSettings array and voila, you have your dynamic settings now loaded into your application.

**4.5. Amazon web service**

The Amazon Web Service API is quite simple and only requires a few lines of code to allow the developer to request a query and receive a collection of items. A sample piece of code to perform a general keyword search on Amazon's catalogs is as below:

```
1: AmazonSearchService amazonSrch = new AmazonSearchService();
2: KeywordRequest kwdReq = new KeywordRequest();
3:
4: kwdReq.devtag = "1010101010";
5: kwdReq.keyword = "Beenie Man";
6: kwdReq.type = "heavy";
7: kwdReq.mode = "music";
8:
9: ProductInfo prodMfg = amazonSrch.KeywordSearchRequest(kwdReq);
```

The three key objects are AmazonSearchService, KeywordRequest, and ProductInfo. The name of the objects alone gives you an idea of their purpose in the whole scheme of things. The AmazonSearchService is what we use to connect to the Web Service offered by Amazon. By this object we can perform various types of searches/requests as provided by the API. One such type search/request is the KeywordRequest which allows us to do a search of the Amazon catalog by simple providing a keyword, quite similar to how one searches a regular web search engine like Google. Each of the various requests offered by the API returns some form of product information regarding the item(s) it found in the catalog. This information encapsulated in the ProductInfo object. The ProductInfo object itself contains a collection of Details objects.

Each time that we get some product information, we make this information available to our DataGrid so that it can be displayed to the end-user during their viewing of the results. We do this by adding the information to our DataSet which will later be bounded to the DataGrid.

```
1: private void AddToDataSet(ProductInfo newProducts)
2: {
3:    DataRow newRow;
4:
5:    for( int i = 0; i < newProducts.Details.Length; i++ )
6:    {
7:       newRow = dsetAmazonReponse.Tables["AmazonResults"].NewRow();
8:
9:       newRow["Our Price"] = newProducts.Details[i].OurPrice;
10:      newRow["List Price"] = newProducts.Details[i].ListPrice;
11:      newRow["Catalog"] = newProducts.Details[i].Catalog;
12:
13:      dsetAmazonReponse.Tables["AmazonResults"].Rows.Add(newRow);
14:   }
15: }
```

As you can see above, every new item is placed in a new row of our DataSet (line 13) and the columns of the row are populated with the actual data from the Details object (line 9-11). At the end, this method allowed us to collect all the product information we need to display to the user.

## 5. CONCLUSION

Ubiquitous, or Pervasive, computing is becoming a fact of life. It gives users an immense amount of freedom and power to do their day-to-day computing tasks. Underneath this power and flexibility visible to end-users is a tremendous amount of software complexity. Consider the various protocols, formats, APIs, platforms, layers, tools, frameworks that come into play when an application delivers its intended functionality. Pulling these elements together to create effective applications that provide the right function and experience to end-users is still a nascent area that is ripe with research challenges.

In this paper, we have presented a new Architecture model for pervasive computing with implementing a new application using Amazon Web service with ASP.NET as programming language. The service API of Amazon Web is quite simple and no hard to use and it is not difficult the utilization of ASP.NET because it provides a powerful functionality in a simple interfacing by the mediator of commands and the elegant components. Pervasive computing applications can be classified in three categories based on interaction models: user-agent to user-agent, user-agent to environment and user-agent to internet). User studies are very important for understanding what users want and are the final measure of progress for pervasive computing. The four key research challenges facing us today are: human-computer interaction, limited battery lifetime on mobile devices, data privacy and spontaneous interaction between devices (via tags, wireless interfaces or video/audio). Our search engine provides an effective form of targeted marketing by creating a personalized shopping experience for each customer. For large retailers like Amazon.com, a good suggestion engine is scalable over very large customer bases and product catalogs, requires only a few seconds processing time to generate online recommendations.